\documentclass[fleqn,usenatbib]{mnras}
\usepackage{graphicx}
\usepackage{float}
\usepackage[utf8]{inputenc}
\usepackage{appendix}
\usepackage{CJK}
\usepackage{multirow}

\usepackage{newtxtext,newtxmath}

\usepackage[T1]{fontenc}

\DeclareRobustCommand{\VAN}[3]{#2}
\let\VANthebibliography\thebibliography
\def\thebibliography{\DeclareRobustCommand{\VAN}[3]{##3}\VANthebibliography}

\usepackage{graphicx}	
\usepackage{amsmath}	
\usepackage{academicons}
\usepackage{xcolor}
\usepackage{orcidlink}

\title{Hessian-based photometric substructure as an evolutionary tracer of OB cluster candidates in M31}

\author[Y. Liang et al.]{
Yuan Liang,$^{1}$\thanks{E-mail: liangyuan@bao.ac.cn}
Chao-Wei Tsai\orcidlink{0000-0002-9390-9672}$^{1,2,3}$\thanks{E-mail: cwtsai@nao.cas.cn},
and Jingwen Wu\orcidlink{0000-0001-7808-3756}$^{1,2}$\thanks{E-mail: jingwen@nao.cas.cn}
\\
$^{1}$School of Astronomy and Space Science, University of Chinese Academy of Sciences, Beijing 100049, People's Republic of China\\
$^{2}$National Astronomical Observatories, Chinese Academy of Sciences, Datun Road A20, Beijing, People's Republic of China\\
$^{3}$Institute for Frontiers in Astronomy and Astrophysics, Beijing Normal University, Beijing 102206, People's Republic of China\\
}

\date{Accepted XXX. Received YYY; in original form ZZZ}

\pubyear{\the\year{}}

\begin{document}
\begin{CJK*}{UTF8}{gbsn}
\label{firstpage}
\pagerange{\pageref{firstpage}--\pageref{lastpage}}
\maketitle

\begin{abstract}
Using \textit{Hubble Space Telescope} images from the PHAT and PHAST surveys, we construct an updated catalogue of 747 OB cluster (OBC) candidates. We introduce a dimensionless structural metric, the trace coefficient of variation ($CV_{\rm tr}$), derived from the Hessian matrix in four \textit{HST} bands, to quantify the internal photometric substructure of partially resolved OBC candidates. Cross-matching with the subset of M31 clusters that have independent colour--magnitude diagram (CMD) age estimates yields 254 objects in common. We find statistically significant anti-correlations between $CV_{\rm tr}$ and age in the UV and blue bands, suggesting a progressive smoothing of the light distribution as clusters evolve. Given the substantial intrinsic scatter, $CV_{\rm tr}$ should not be interpreted as a precise age indicator for individual clusters, but rather as a statistical structural proxy for ensemble cluster populations. Bootstrap resampling confirms the robustness of these trends. Forward modelling of synthetic clusters analysed with the same pipeline reproduces an overall decline of $CV_{\rm tr}$ with age, while also revealing wavelength-dependent departures from a simple monotonic relation at young ages. These results show that second-order photometric structure contains measurable evolutionary information within the CMD-calibrated regime ($\sim10$--300~Myr).
\end{abstract}




\section{Introduction}\label{sec:intro}

Star clusters are gravitationally bound stellar systems formed in nearly coeval star-forming events, and serve as fundamental laboratories for studying star formation, stellar feedback, stellar evolution, and galactic dynamics \citep{krumholz2019star, krause2020physics}. In particular, the early evolution of star clusters remains an open problem. Despite extensive observational and theoretical efforts, the dynamical state and long-term survival of young stellar systems are still uncertain \citep{adamo2020star}.

In the Milky Way, the proximity of young star clusters allows individual stars to be resolved, enabling detailed analyses based on colour--magnitude diagrams (CMDs) and tailored $N$-body modelling \citep{allen2007structure, jaehnig2015structural, kuhn2015spatial, zhong2022new, tarricq2022structural}. These approaches provide some of the most direct constraints on cluster ages, structures, and dynamical states. By contrast, star clusters located in galaxies beyond several megaparsecs provide comparatively little information on their internal structure, and are therefore typically treated as integrated light sources. Their physical properties are commonly inferred from simplified morphological, photometric, or spectral models \citep{king1966structure, elson1987structure, barmby2009hubble, ryon2017effective, brown2021radii, maschmann2024phangs, thilker2025phangs}.

With the increasing spatial resolution of modern imaging surveys, star clusters in nearby galaxies are now observed in an intermediate regime where most individual stars remain unresolved, but internal structural features are clearly detectable. In this regime, classical CMD analyses are not applicable to the majority of compact or crowded clusters, while treating these systems as simple or spherically symmetric objects would result in the loss of valuable internal structural information. Consequently, more suitable methods are required to analyse partially resolved star clusters and to fully exploit the available structural information.

A number of recent studies have already begun to explore such approaches. One example is the pixel colour--magnitude diagram (pCMD) technique, which exploits pixel-level flux distributions to extract statistical information from partially resolved stellar populations \citep{cook2019measuring, cook2020measuring}. \citet{whitmore2011using} used H$\alpha$ morphology and pixel-to-pixel surface-brightness fluctuations to constrain the ages of young star clusters in M83. In parallel, large surveys such as LEGUS and PHANGS have investigated young star clusters by treating them as extended two-dimensional objects, incorporating structural and morphological measurements derived from high-resolution imaging \citep{adamo2017legacy, cook2019star, wei2020deep, hannon2023star}. While these studies demonstrate the potential of moving beyond purely integrated descriptions, they primarily rely on zeroth- or first-order structural information and do not fully exploit higher-order features associated with internal substructure.

To address this challenge, we introduce a Hessian-based framework for analysing the internal photometric structure of partially resolved star clusters. The Hessian matrix, a second-order differential operator sensitive to curvature and filamentary structure, has been widely applied in image processing \citep{lowe2004distinctive, xiao2019multi}, oceanography \citep{loose2021leveraging}, seismology \citep{yang2025high}, quantum physics \citep{dawid2021hessian}, and medical imaging \citep{yang2014fast, lavin2024improved}. In astronomy, however, this approach has mainly been used in studies of the cosmic web and large-scale structure \citep{pfeifer2022cows, olex2025universal}.

The performance of the Hessian-based framework depends sensitively on image resolution: insufficient spatial sampling suppresses detectable substructure, while excessively large image sizes increase computational cost without proportionate gains in structural information. For young clusters with typical half-light radii of a few parsecs \citep{brown2021radii}, the clusters typically have radii of order a few to a few tens of pixels (at $\sim$0.04--0.05\arcsec\,pixel$^{-1}$) at distances of 1--3 Mpc. Consequently, the Andromeda Galaxy (M31), the nearest large spiral galaxy, provides an ideal data set for developing and testing our framework. It hosts a wide variety of star clusters that span a range of ages and masses, with varying resolutions in available imaging data. Some of these clusters are available for colour-magnitude diagram age determination, which can be cross-validated with the structural analyses provided by our Hessian-based method.

As our goal is to investigate the early evolutionary stages of star clusters, careful sample selection is essential. Young clusters dominated by O- and B-type stars provide particularly valuable laboratories in this context, owing to their short lifetimes, high masses, and strong stellar feedback. Over the past decade, the PHAT \citep{dalcanton2012panchromatic, williams2014panchromatic, williams2023panchromatic} and PHAST \citep{chen2025phast} surveys have delivered extensive multi-band imaging of M31, enabling detailed cataloguing and analysis of cluster properties, including ages, masses \citep{beerman2012panchromatic, fouesneau2014panchromatic}, and formation efficiencies \citep{johnson2016panchromatic, johnson2017panchromatic}, as well as the construction of homogeneous cluster catalogues \citep{johnson2012phat, johnson2015phat}. Despite these advances, most existing analyses rely on integrated light measurements or global structural parameters, and thus largely overlook the internal structural complexity of young or morphologically irregular clusters, particularly in crowded environments \citep{de2017deriving, naujalis2021deriving, krivsvciunas2023deriving}.

In our previous work \citep{liang2025comprehensive}, we identified a population of OB cluster (OBC) candidates in M31 based on their compact morphology and UV-bright stellar content. In the present study, we first extend that catalogue by adopting improved selection criteria and cross-match our catalogue with the catalogue of \citet{johnson2016panchromatic}, which provides CMD-based ages for a resolvable subset of M31 clusters. We then apply the Hessian-derived trace coefficient of variation ($CV_{\rm tr}$) to quantify the internal photometric substructure of OBC candidates. The analysis reveals statistically significant correlations between $CV_{\rm tr}$ and cluster age within the CMD-calibrated regime. Finally, to investigate the physical plausibility of the observed trends, we construct synthetic cluster models spanning a broad range of ages and process them through the same measurement pipeline. This forward-modelling approach allows us to evaluate whether the empirical $CV_{\rm tr}$--age relation can naturally arise from simplified but physically motivated assumptions about stellar population evolution and structural fading.

The paper is organized as follows. In Section~\ref{sec:data_method}, we describe the observational datasets, present the algorithms for improving the OBC candidate catalogue, and introduce the Hessian-based structural indicators. Section~\ref{sec:results} shows the updated catalogue and the spatial and evolutionary trends of the structure indices. In Section~\ref{sec:discussion}, we evaluate the catalogue robustness and assess the performance of the Hessian-based indicators. Section~\ref{sec:conclusion} summarizes our findings and outlines future directions.

\section{Data and Methodology} \label{sec:data_method}

\subsection{Data} \label{sec:data}

This study is based on imaging from the Panchromatic Hubble Andromeda Treasury (PHAT; \citealt{dalcanton2012panchromatic}) and the Panchromatic Hubble Andromeda Southern Treasury (PHAST; \citealt{chen2025phast}) surveys. Together, these programmes provide near-contiguous, high-resolution coverage of approximately two-thirds of the M31 disc, extending from the central bulge to galactocentric radii of $\sim$20\,kpc.

The PHAT survey covers $\sim$0.5\,deg$^{2}$ in the north-eastern disc using 828 \textit{HST} orbits, while PHAST expands the footprint by an additional $\sim$0.45\,deg$^{2}$ along the southern major axis. Both programmes include imaging in four broadband filters (F275W, F336W, F475W, and F814W) obtained with ACS and WFC3. Observing strategies were coordinated to ensure consistent spatial resolution ($\sim$0.05\,arcsec\,pixel$^{-1}$ in the optical bands) and photometric depth, with Nyquist-sampled imaging in F475W and F814W to enable accurate reconstruction of point sources.

The combined dataset resolves more than $1.5\times10^{8}$ stars across ultraviolet, optical, and near-infrared wavelengths \citep{dalcanton2012panchromatic, williams2014panchromatic, williams2023panchromatic, chen2025phast}. Photometric completeness in the F475W band reaches $\sim$27.9\,mag in the outer disc and declines to $\sim$26.0\,mag in more crowded inner regions. Completeness and photometric uncertainties were quantified using artificial star tests and crowding simulations; detailed procedures are described in \citet{dalcanton2012panchromatic} and \citet{chen2025phast}.

\subsection{Identification via MeanShift and Photometric Measurement} \label{sec:meanshift}

Young cluster candidates are initially identified using F275W imaging, which is particularly sensitive to the strong ultraviolet emission from massive stars. Longer-wavelength bands (F336W, F475W, and F814W) are subsequently used for morphological verification, background rejection, and structural characterization. Our identification and photometric measurement procedures follow \citet{liang2025comprehensive}, with two updates: (i) an extended set of spatial scales for the MeanShift search and (ii) a semi-automated estimate of the full-light radius, $R_{\rm full}$. We summarize the key steps of our method below and present the resulting catalogue of OBC candidates in Section \ref{sec:extend_cat} (Table \ref{tab:obc_cat}).

Candidate clusters were identified using the F275W images from the \textit{HST} PHAT and PHAST surveys, publicly available via MAST.\footnote{\url{https://archive.stsci.edu/hlsp/phat}; \url{https://archive.stsci.edu/hlsp/phast}} The images were processed with the standard \textit{HST} pipelines. To enhance sensitivity to local overdensities on a broader range of physical scales, we adopt a set of MeanShift search radii spanning 25--200 pixels, extending the 25--150 pixel range used in \citet{liang2025comprehensive}.

We applied the MeanShift algorithm \citep{fukunaga1975estimation, comaniciu1997robust, bradski1998real, comaniciu1999mean, li2009convergent} to identify density peaks. Each candidate overdensity is required to be associated with at least five detected sources, where sources are defined as local intensity maxima identified with \texttt{Photutils} \citep{Larry2024photutils}. This detection strategy does not require sources to be fully resolved or isolated, which is advantageous in the partially resolved regime. Foreground contaminants were removed via cross-matching to Gaia DR3 \citep{brown2021gaiadr3, evans2023gaia}, and all candidates were subsequently verified by visual inspection.

To determine the photometric aperture, we developed a semi-automated procedure to estimate the full-light radius, $R_{\rm full}$, improving consistency relative to manual measurements. The cluster centre is given by the MeanShift peak, and we extract a $4''$ cutout motivated by the typical M31 cluster size distribution \citep{liang2025comprehensive}. We compute cumulative fluxes in concentric circular apertures and define $R_{\rm full}$ as the smallest radius at which the growth curve converges, i.e.,
$F(r)/F(r+1) > 0.98$ for two consecutive steps, where $F(r)$ is the cumulative flux enclosed within radius $r$ sampled in one-pixel steps. We further impose an upper limit of $R_{\rm full} < 2''$ to prevent noise-driven divergence of the growth curve in crowded or low surface-brightness cases. While empirical, this criterion provides a practical and reproducible definition of cluster extent for irregular morphologies.

The half-light radius, $R_{\rm eff}$, is estimated using two parametric fits: a King profile \citep{king1966structure} and an EFF profile \citep{elson1987structure}, which are commonly used to describe bound and extended clusters, respectively. When both fits are successful, the adopted $R_{\rm eff}$ is taken as the average of the two estimates. If either model fails to yield a valid fit, the $R_{\rm eff}$ derived from the other model is adopted. For every cluster in our sample, at least one of the two models yielded a valid $R_{\rm eff}$.

Photometry is performed within $R_{\rm full}$. Local backgrounds are estimated from annuli spanning $1.2\,R_{\rm full}$ to $3.4\,R_{\rm full}$. Aperture corrections for flux beyond $R_{\rm full}$ assume a King profile with $c=7$ \citep{liang2025comprehensive}. The total flux is computed either using a direct $R_{\rm total}$ measurement where feasible or using the empirical scaling $R_{\rm total}=2.5\,R_{\rm full}$. Magnitudes are calibrated using the standard WFC3/UVIS zeropoints. All measurements are visually inspected to exclude spurious sources and instrumental artefacts.

Uncertainties in $R_{\rm eff}$, $R_{\rm full}$, and the integrated apparent magnitudes were estimated by propagating the corresponding measurement uncertainties. The local background level and root-mean-square (rms) noise were determined using sigma-clipped statistics.

For $R_{\rm full}$, the background noise was propagated to the cumulative aperture fluxes and subsequently to the adjacent flux ratios. The formal $1\sigma$ uncertainty in $R_{\rm full}$ was then estimated by propagating the uncertainty in the adjacent flux ratio to the threshold-crossing radius using the local radial gradient of the flux-ratio profile.

For $R_{\rm eff}$, uncertainties were evaluated independently for the EFF and King profile measurements by propagating the fitted-parameter covariance matrices to the corresponding half-light radii. When both measurements were available, the statistical uncertainty of $R_{\rm eff}$ was conservatively estimated as
$\sigma_{\rm stat}=(\sigma_{\rm EFF}+\sigma_{\rm King})/2$,
while the model-dependent uncertainty was taken as
$\sigma_{\rm model}=|R_{\rm eff,EFF}-R_{\rm eff,King}|/2$.
The final uncertainty was then calculated as
$\sigma_{R_{\rm eff}}=(\sigma_{\rm stat}^{2}+\sigma_{\rm model}^{2})^{1/2}$. When only one profile yielded a valid measurement, its half-light radius and propagated uncertainty were adopted.

For the integrated apparent magnitudes, the photometric uncertainties were estimated from the signal-to-noise ratios of the aperture-photometry measurements and combined in quadrature with the photometric zero-point uncertainty.

Both $R_{\rm full}$ and $R_{\rm eff}$ are determined from the F275W image and are not re-derived independently in the other bandpasses. The F275W-defined spatial region is retained for the subsequent multi-band analysis, ensuring that the structural measurements in different filters sample the same physical region of each OBC candidate.

\subsection{Structural Curvature from the Hessian Matrix}\label{sec:hessian}

Higher-order spatial derivatives of an image, such as curvature, are intrinsically sensitive to fine-scale variations in surface brightness and therefore provide a useful probe of internal structural complexity. In young star clusters, the presence of luminous massive O- and B-type stars is expected to produce pronounced local curvature variations, particularly in the ultraviolet (UV), where these stars contribute strongly to the integrated light. As clusters evolve, other luminous stellar populations may become prominent at different wavelengths and produce similar localized curvature features. In particular, red supergiants (RSGs), the evolved counterparts of massive main-sequence stars, and luminous asymptotic giant branch (AGB) stars may contribute significantly to such structures at redder wavelengths.

Motivated by these considerations, we adopt curvature as a quantitative structural metric to characterize and compare the evolutionary states of star clusters. In practice, curvature is measured through the trace of the Hessian matrix, $\mathrm{Tr}(\mathbf{H})$, which captures the local isotropic curvature of the surface-brightness distribution and is sensitive to rapid intensity variations associated with compact cores, extended haloes, and embedded substructures. We note that the absolute normalization of the curvature statistic is band-dependent due to differing pixel scales and noise properties; our interpretation therefore focuses on monotonic trends within each band rather than on cross-band absolute values.

The photometric image of a star cluster is represented as a two-dimensional scalar field $I(x,y)$, from which the Hessian matrix is defined as
\begin{equation}
\mathbf{H}(x, y) =
\begin{bmatrix}
\frac{\partial^2 I}{\partial x^2} & \frac{\partial^2 I}{\partial x \partial y} \\
\frac{\partial^2 I}{\partial y \partial x} & \frac{\partial^2 I}{\partial y^2}
\end{bmatrix}.
\end{equation}

Second-order derivatives amplify pixel-scale noise and are therefore evaluated on a smoothed version of the image. Each cutout is first convolved with a Gaussian kernel. To preserve a consistent angular smoothing scale across bands with different pixel sizes, we adopt $\sigma_{\rm G}=2$ pixels for the UV bands (pixel scale $0.0396\arcsec\,{\rm pixel}^{-1}$) and $\sigma_{\rm G}=2.5$ pixels for the F475W/F814W bands (pixel scale $0.05\arcsec\,{\rm pixel}^{-1}$). These values correspond to the same angular smoothing scale within the two pixel samplings.

Second-order derivatives are evaluated on the smoothed image using standard three-point central finite-difference operators in pixel coordinates. Specifically, after Gaussian smoothing, we compute $\partial^2 I/\partial x^2$, $\partial^2 I/\partial y^2$, and $\partial^2 I/\partial x\partial y$ via discrete central differences, yielding pixel-wise estimates of the Hessian components and hence $\mathrm{Tr}(\mathbf{H})$.

To suppress low-level background fluctuations while retaining pixels associated with relatively strong curvature features, we define an effective region $\Omega_{\mathrm{eff}}$ as
\begin{equation}
\Omega_{\mathrm{eff}} \equiv \left\{(x,y): 
\left|\mathrm{Tr}\left[\mathbf{H}(x,y)\right]\right| >
\mathrm{median}\left(\left|\mathrm{Tr}\left[\mathbf{H}\right]\right|\right)
\right\}.
\end{equation}
That is, $\Omega_{\mathrm{eff}}$ consists of pixels whose absolute curvature exceeds the median value within the cutout. Unlike a surface-brightness selection, $\Omega_{\mathrm{eff}}$ is sensitive to local intensity variations, such that bright but spatially smooth regions may be excluded, whereas fainter but sharply varying features may satisfy the curvature-based selection criterion.

Scaling the curvature threshold by factors of 0.7--1.3 results in systematic but mild changes in $CV_{\rm tr}$, typically at the 6--15\% level across representative clusters. The variation is monotonic and smooth, reflecting the progressively more restrictive selection of high-curvature pixels. No abrupt transitions or qualitative changes are observed, indicating that the metric is only weakly sensitive to moderate changes in the threshold definition.

The trace coefficient of variation is defined as the coefficient of variation of the Hessian-trace amplitude within $\Omega_{\mathrm{eff}}$:
\begin{equation}
CV_{\mathrm{tr}} = 
\frac{\sigma_{|\mathrm{Tr}|}}{\mu_{|\mathrm{Tr}|}},
\end{equation}
where $\mu_{|\mathrm{Tr}|}$ and $\sigma_{|\mathrm{Tr}|}$ are the mean and standard deviation of $|\mathrm{Tr}(\mathbf{H})|$ evaluated over $\Omega_{\mathrm{eff}}$. Because $CV_{\rm tr}$ is a coefficient of variation, it is invariant under an overall multiplicative rescaling of the curvature amplitude and therefore primarily reflects the relative heterogeneity of curvature values within the selected region. Higher values of $CV_{\mathrm{tr}}$ correspond to stronger spatial variations in local curvature and hence to increased internal structural complexity.

\begin{figure}
    \centering
    \includegraphics[width=1 \linewidth]{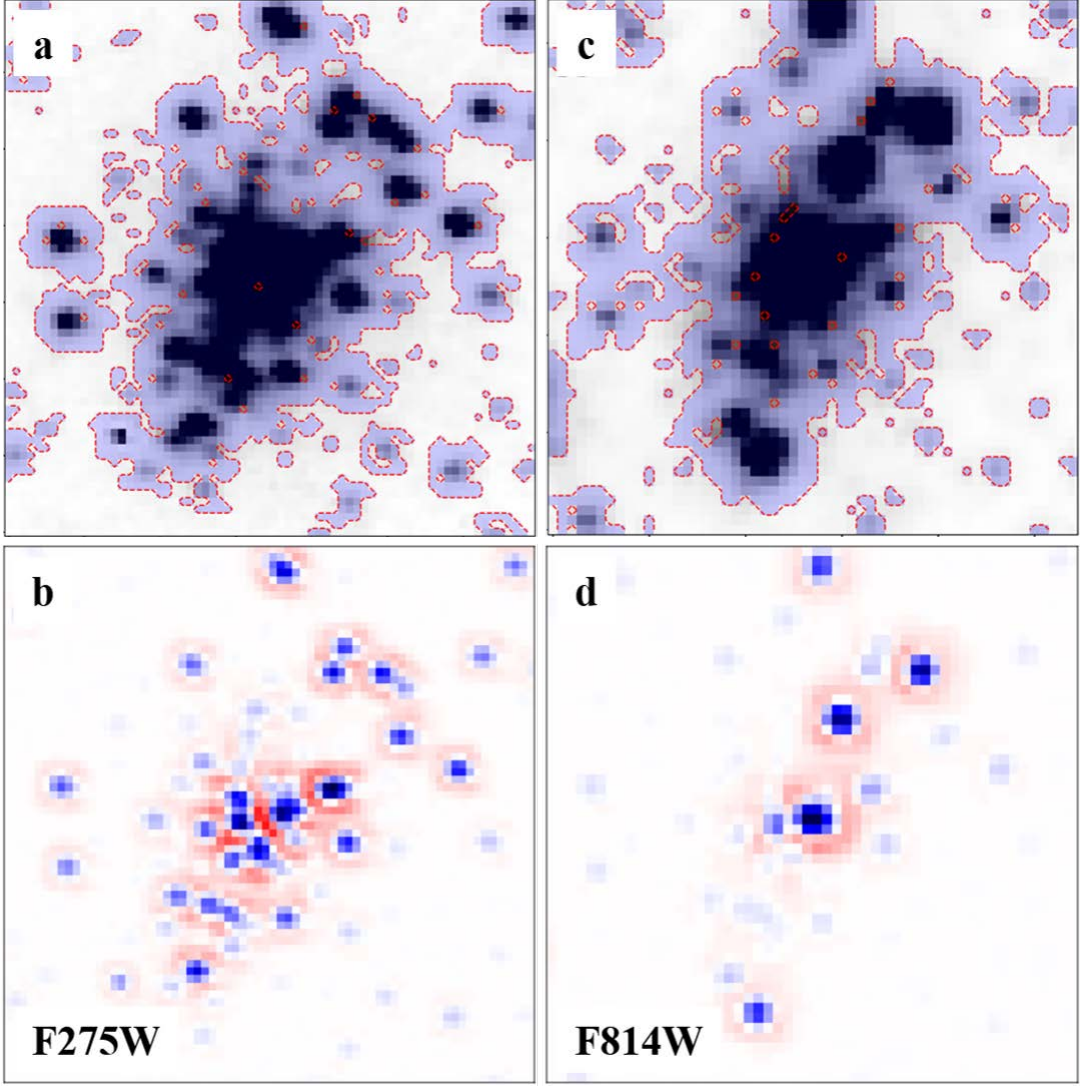}
    \caption{
    Example of Hessian-based curvature analysis applied to a star cluster in two photometric bands.
    Panels \textbf{(a)} and \textbf{(b)} show the surface brightness map and corresponding curvature (Hessian trace) map in the F275W band, respectively, while panels \textbf{(c)} and \textbf{(d)} show the same quantities in the F814W band.
    In panels \textbf{(a)} and \textbf{(c)}, the pixels belonging to the effective region $\Omega_{\rm eff}$ are highlighted by semi-transparent light-blue shading, while the red dashed contours trace the pixel-based boundaries of the selected regions.}
    In the curvature maps, blue regions indicate locally convex intensity distributions, while red regions indicate concave structures.
    \label{fig:Hessian}
\end{figure}

Figure~\ref{fig:Hessian} illustrates the Hessian-based curvature analysis for an example OBC candidate in two photometric bands. In the surface-brightness maps shown in panels \textbf{(a)} and \textbf{(c)}, the light-blue shading marks the pixels included in the effective region $\Omega_{\rm eff}$, while the red dashed contours delineate their pixel-based boundaries. The selected pixels need not form a single contiguous region, but can instead comprise multiple disconnected regions of different sizes, reflecting the local spatial structure of the light distribution. Although the surface-brightness morphologies appear broadly similar, the curvature maps reveal pronounced differences in the spatial distribution and amplitude of local structural variations. While only two bands are shown for clarity, the curvature analysis is performed consistently across all four photometric bands considered in this work, allowing complementary structural information to be extracted at different wavelengths.

\section{Results} \label{sec:results}

\subsection{The Extended Catalogue of OBC candidates in M31} \label{sec:extend_cat}

\begin{figure}
    \centering
    \includegraphics[width=1 \linewidth]{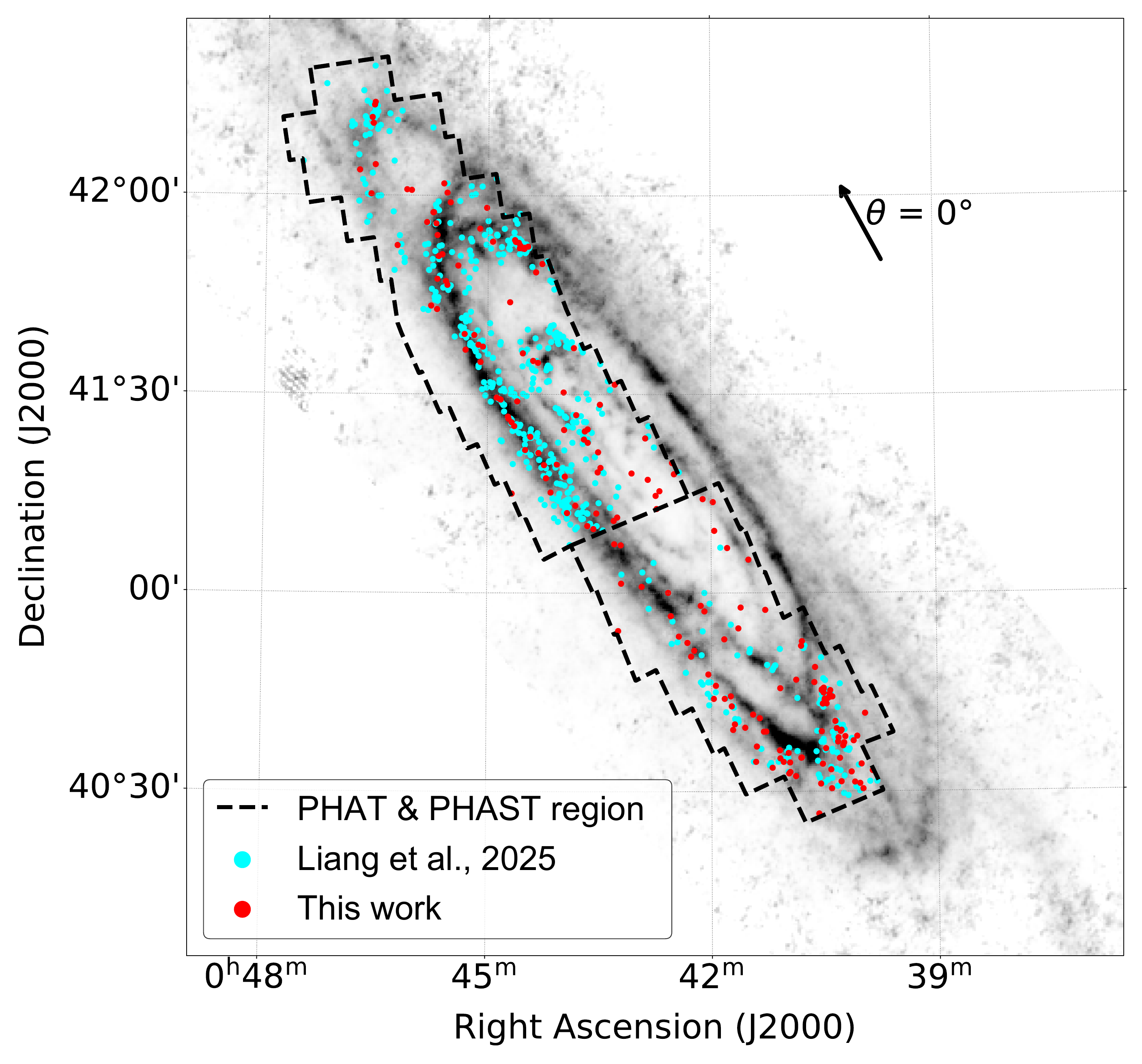}
    \caption{Spatial distribution of candidate OBCs in M31, overlaid on the dust emission map from \citet{draine2013andromeda}. Cyan dots represent the OBC candidates from \citet{liang2025comprehensive}, and red dots are newly identified candidates from this work. The dashed box outlines the combined PHAT/PHAST survey footprint. The boundary between the PHAT (north) and PHAST (south) survey regions is marked by the central dashed line. A black arrow indicates the projected north-eastern major axis of M31, which defines the reference direction ($\theta = 0^\circ$) for the position angle relative to the projected major axis, used in subsequent analyses.
    }
    \label{fig:OBC_layout}
\end{figure}

Fig.~\ref{fig:OBC_layout} shows the spatial distribution of OBC candidates across M31, overlaid on the galaxy's dust emission map from \citet{draine2013andromeda}. The grayscale background highlights the intensity of the emission. Cyan points denote candidates identified in our previous study \citep{liang2025comprehensive}, while red points mark new detections from the current work. The PHAT/PHAST boundary is indicated by the central dashed line, while the dashed box marks the combined survey footprint. The newly identified candidates broadly trace the morphology of the galactic disc and are spatially consistent with previously identified OBCs \citep{liang2025comprehensive}. With the refined grid-search configuration described in Section~\ref{sec:meanshift}, the total number of candidates increases from 578 to 747. The increase is mainly due to the larger window sizes in our refined MeanShift configuration (Section~\ref{sec:meanshift}), which allow more extensive scale searches. These larger windows make it possible to detect candidates in crowded regions, such as the southern part of M31, where smaller windows may have treated them as background stars. The newly identified candidates in these areas are generally larger than previously detected southern clusters, while showing similar brightness.

\begin{figure}
    \centering
    \includegraphics[width=1\linewidth]{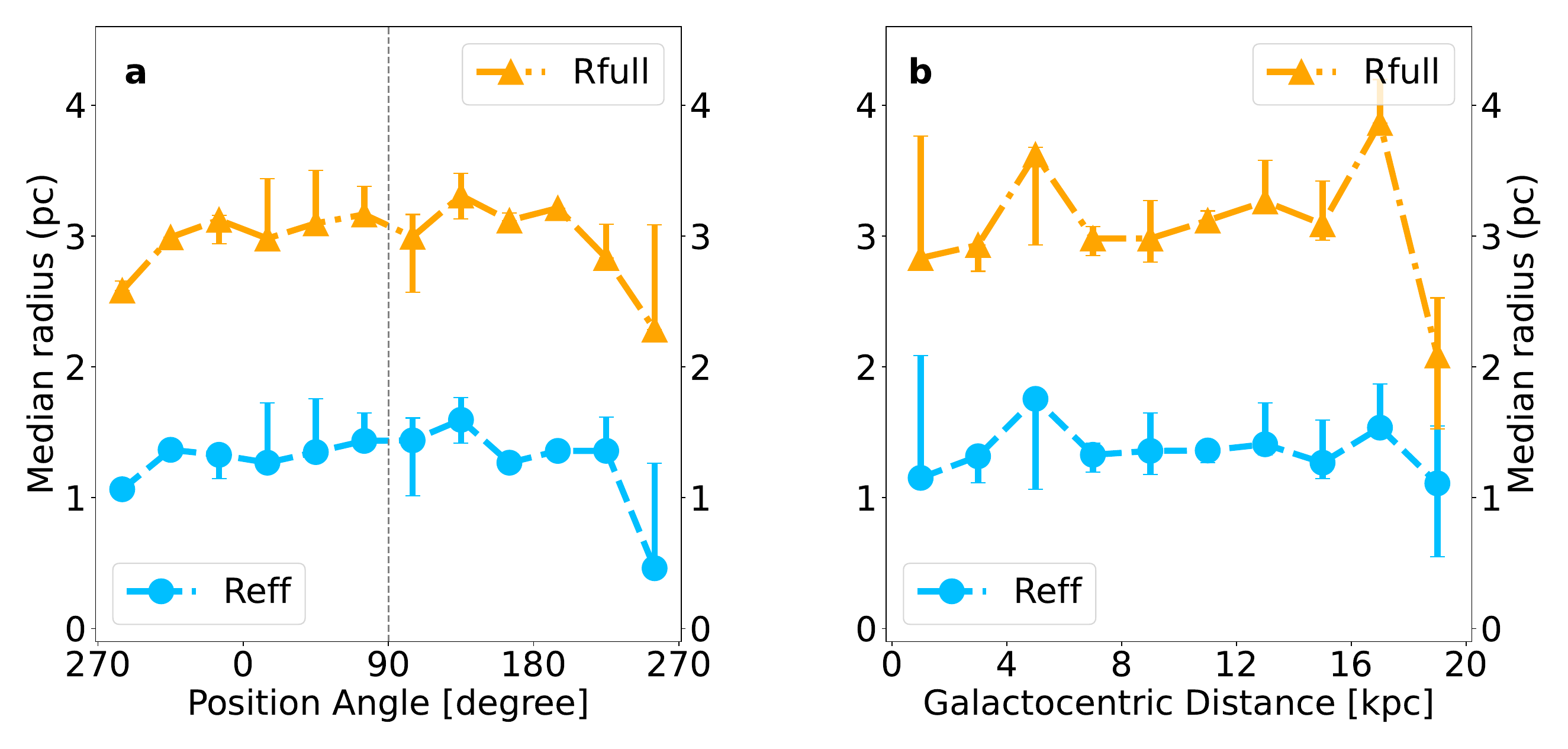}
    \caption{
    Spatial variation of the characteristic sizes of OBC candidates in the F275W band.
    (\textbf{a}) Median half-light radius ($R_{\rm eff}$; blue circles, dashed line) and full-light radius ($R_{\rm full}$; orange triangles, dot-dashed line) as a function of position angle, measured counterclockwise from the projected northern major axis of M31.
    The vertical dashed line at $\theta = 90^\circ$ separates the northern and southern regions.
    (\textbf{b}) The same radii plotted against projected galactocentric distance.
    In both panels, the y-axis shows the median radius in parsecs (pc)}. Error bars indicate the 25th--75th percentile range within each bin.
    \label{fig:reff}
\end{figure}

Fig.~\ref{fig:reff} shows the spatial dependence of the characteristic OBC sizes in the F275W band, quantified by the half-light radius ($R_{\rm eff}$) and full-light radius ($R_{\rm full}$). These are shown as functions of position angle (Fig.~\ref{fig:reff}a) and projected galactocentric distance (Fig.~\ref{fig:reff}b). The position angle is measured counterclockwise from the projected northern major axis of M31, corresponding to $\theta = 0^\circ$ as illustrated in Fig.~\ref{fig:OBC_layout}.

In Fig.~\ref{fig:reff}a, OBC candidates in the northern region (left of the dashed line) exhibit slightly larger median $R_{\rm eff}$ values and broadly comparable $R_{\rm full}$, consistent with our earlier findings \citep{liang2025comprehensive}. However, the application of the updated size estimation procedure yields systematically smaller size measurements compared to those reported in our previous work. This mainly reflects the more conservative, growth-curve-based definition of $R_{\rm full}$ adopted here, which reduces the inclusion of low-level diffuse light in crowded backgrounds. As a result, the $R_{\rm full}$ values decrease from $\sim$8--10\,pc in the north and 3--6\,pc in the south to a converged median of $\sim$3.1\,pc for the combined sample. The updated $R_{\rm eff}$ values yield median sizes of 1.29\,pc and 1.35\,pc in the north and south, respectively. These values are consistent with the typical half-light radii of young star clusters reported in nearby galaxies, which are commonly found to be of order 1--2\,pc \citep[e.g.][]{brown2021radii}.

By contrast, Fig.~\ref{fig:reff}b shows no clear systematic trend with galactocentric distance, aside from a decrease in the outermost bin ($R > 18$\,kpc). This may be due to the small number of sources ($N=4$), or it may reflect differences in the underlying stellar populations at large radii, which tend to appear more compact at UV wavelengths. In addition, both $R_{\rm eff}$ and $R_{\rm full}$ show a local increase near a galactocentric distance of $\sim5$ kpc; however, the interquartile ranges overlap substantially with those of the neighbouring bins, and we do not identify a clear physical origin for this feature.

\begin{table*}
\centering
\caption{The extended catalogue of OBC candidates in M31.}
\label{tab:obc_cat}
\small
\setlength{\tabcolsep}{4pt}

{\begin{tabular}{llccrrrrcrrrcc}
\hline
YLID & Brick & R.A. & Decl. &
$R_{\rm full}$ & $\sigma_{R_{\rm full}}$ &
$R_{\rm eff}$ & $\sigma_{R_{\rm eff}}$ &
Filter & $m$ & $\sigma_m$ & $CV_{\rm tr}$ &
APID & Age \\
 & & (deg) & (deg) &
($^{\prime\prime}$) & ($^{\prime\prime}$) &
(pc) & (pc) &
 & (mag) & (mag) & &
 & (Myr) \\
\hline

\multirow{4}{*}{1}
& \multirow{4}{*}{b01}
& \multirow{4}{*}{10.838138}
& \multirow{4}{*}{41.214752}
& \multirow{4}{*}{0.87}
& \multirow{4}{*}{0.03}
& \multirow{4}{*}{1.81}
& \multirow{4}{*}{0.06}
& F275W & 22.09 & 0.47 & 1.19
& \multirow{4}{*}{--}
& \multirow{4}{*}{--} \\
&&&&&&&& F336W & 21.51 & 0.79 & 1.15 && \\
&&&&&&&& F475W & 20.46 & 0.78 & 0.86 && \\
&&&&&&&& F814W & 20.12 & 0.64 & 1.05 && \\[2pt]

\multirow{4}{*}{2}
& \multirow{4}{*}{b01}
& \multirow{4}{*}{10.817428}
& \multirow{4}{*}{41.242983}
& \multirow{4}{*}{0.63}
& \multirow{4}{*}{0.02}
& \multirow{4}{*}{1.37}
& \multirow{4}{*}{0.02}
& F275W & 23.20 & 0.53 & 0.59
& \multirow{4}{*}{--}
& \multirow{4}{*}{--} \\
&&&&&&&& F336W & 22.35 & 0.99 & 0.64 && \\
&&&&&&&& F475W & 20.60 & 0.65 & 0.61 && \\
&&&&&&&& F814W & 19.70 & 0.33 & 0.60 && \\[2pt]

\multirow{4}{*}{4}
& \multirow{4}{*}{b01}
& \multirow{4}{*}{10.607419}
& \multirow{4}{*}{41.303992}
& \multirow{4}{*}{0.71}
& \multirow{4}{*}{0.02}
& \multirow{4}{*}{0.24}
& \multirow{4}{*}{0.01}
& F275W & 21.59 & 0.27 & 2.17
& \multirow{4}{*}{--}
& \multirow{4}{*}{--} \\
&&&&&&&& F336W & 21.23 & 0.43 & 2.32 && \\
&&&&&&&& F475W & 20.84 & 0.91 & 1.19 && \\
&&&&&&&& F814W & 20.39 & 0.70 & 0.56 && \\[2pt]

& \vdots & \vdots & \vdots & \vdots & \vdots & \vdots & \vdots & \vdots & \vdots & \vdots & \vdots && \\

\multirow{4}{*}{12}
& \multirow{4}{*}{b02}
& \multirow{4}{*}{10.927740}
& \multirow{4}{*}{41.218019}
& \multirow{4}{*}{0.44}
& \multirow{4}{*}{0.10}
& \multirow{4}{*}{0.74}
& \multirow{4}{*}{0.03}
& F275W & 22.67 & 0.24 & 0.69
& \multirow{4}{*}{819}
& \multirow{4}{*}{7.8} \\
&&&&&&&& F336W & 22.06 & 0.24 & 0.71 && \\
&&&&&&&& F475W & 20.86 & 0.34 & 0.67 && \\
&&&&&&&& F814W & 20.35 & 0.40 & 0.97 && \\ [2pt]

& \vdots & \vdots & \vdots & \vdots & \vdots & \vdots & \vdots & \vdots & \vdots & \vdots & \vdots && \\

\hline
\end{tabular}
}
\begin{flushleft}
\textit{Notes.} Column ``YLID'' assigns a unique identifier to each candidate, which has been reassigned independently of any previous catalogue. ``Brick'' refers to the PHAT/PHAST survey region where the object is located. $R_{\rm full}$ and $R_{\rm eff}$ represent the full-light radius (in arcseconds) and the effective (half-light) radius (in parsecs), respectively, while their corresponding uncertainties are denoted by $\sigma_{R_{\rm full}}$ and $\sigma_{R_{\rm eff}}$. The apparent magnitude $m$, its uncertainty $\sigma_m$, and the concentration-related parameter $CV_{\rm tr}$ are given for each of the HST F275W, F336W, F475W, and F814W bands. ``APID'' denotes the identifier of the matched source in the Andromeda Project catalogue, and ``Age'' gives the corresponding age in Myr. The complete version of this table is available online at \url{https://nadc.china-vo.org/res/r101747/}.
\end{flushleft}
\end{table*}

We compile all OBC candidates into a catalogue, with a representative subset shown in Table~\ref{tab:obc_cat}. Each entry includes a unique identifier (YLID), the corresponding PHAT/PHAST survey brick, equatorial coordinates (J2000), $R_{\rm full}$ and its uncertainty (in arcseconds), $R_{\rm eff}$ and its uncertainty (in parsecs), the apparent magnitude and its uncertainty in each of the four filters, together with the corresponding $CV_{\rm tr}$ values. The APID and age from the AP cluster catalogue of \citet{johnson2016panchromatic}, obtained by cross-matching with our sample, are also provided. The methods used to derive these quantities are described in Sec.~\ref{sec:meanshift}. YLID values have been renumbered in this work and do not match those in previous catalogues. The $R_{\rm full}$ values listed in Table~\ref{tab:obc_cat} are given in angular units (arcseconds). For the discussion in this section, angular radii are converted to physical units assuming a distance of 0.78\,Mpc for M31. Overall, the extended catalogue preserves the large-scale spatial distribution of OBC candidates reported in previous work, while yielding systematically revised structural parameters under the updated measurement scheme.

\subsection{Trace Coefficient of Variation vs. Ages of clusters}
\label{sec:tensor_time}

Most star clusters at 1--3 Mpc remain unresolved, limiting CMD-based age dating for the majority of compact or crowded systems. We therefore cross-match our extended catalogue with the subset of M31 clusters for which CMD-based ages are available from \citet{johnson2016panchromatic}, and use these ages as an external calibration to test whether $CV_{\rm tr}$ carries evolutionary information. In total, we identify 254 clusters in common with the catalogue of \citet{johnson2016panchromatic}. The cross-match is based on the cluster coordinates. Specifically, we compared the centre position of each cluster in our catalogue with those of the Andromeda Project clusters and considered two sources to be the same object when their angular separation was smaller than 0.36 arcsec. All accepted matches have a unique counterpart within the adopted matching radius.

\begin{figure}
    \centering
    \includegraphics[width=1 \linewidth]{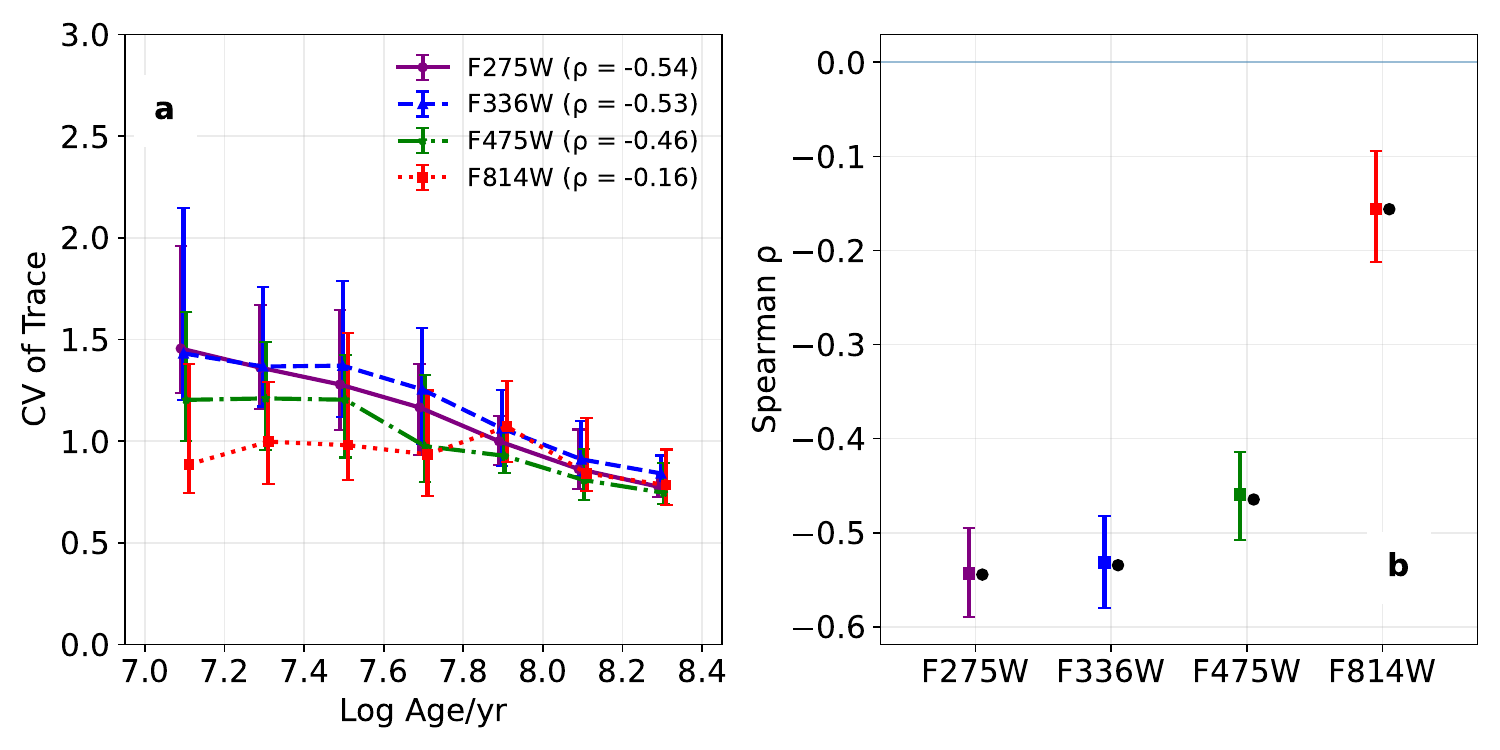}
    \caption{
    Evolution of the Hessian-based structure index $CV_{\rm tr}$ for OB cluster candidates in M31. 
    (\textbf{a}): Median $CV_{\mathrm{tr}}$ as a function of CMD age for 254 clusters. The horizontal axis shows stellar age on a logarithmic scale, spanning $\sim10$~Myr to 300~Myr, with a bin width of 0.2~dex. Coloured symbols and line styles denote the four photometric bands: F275W (purple solid circles), F336W (blue dashed triangles), F475W (green dash-dotted asterisks), and F814W (red dotted squares). Small horizontal offsets are applied for visual separation and do not affect the statistical binning. Vertical bars represent the interquartile range (25th--75th percentile) within each age bin. The Spearman rank correlation coefficients derived from the unbinned data are indicated in the panel. 
    (\textbf{b}): Bootstrap stability test of the Spearman correlation coefficients $\rho$. Squares denote the bootstrap medians of $\rho$ from 1000 resamplings, with error bars corresponding to the 16th--84th percentile range. Black circles indicate the original $\rho$ values computed from the full sample.
    }
    \label{fig:tr_veri_cmd}
\end{figure}

Figure~\ref{fig:tr_veri_cmd}a shows the relation between $CV_{\mathrm{tr}}$ and CMD age in four bands, with interquartile ranges computed within 0.2~dex age bins. A systematic decline of median $CV_{\mathrm{tr}}$ with increasing age is evident in the near-UV and blue bands (F275W, F336W, and F475W), indicating that the UV/blue light distribution becomes progressively smoother and less dominated by high-contrast substructure as clusters evolve.

The Spearman rank correlation coefficients derived from the unbinned data are $\rho = -0.54$ (F275W), $-0.53$ (F336W), $-0.46$ (F475W), and $-0.16$ (F814W), with corresponding $p$-values of $5.3\times10^{-21}$, $3.6\times10^{-20}$, $5.2\times10^{-15}$, and $1.3\times10^{-2}$, respectively. These values indicate a moderately strong and highly significant anti-correlation between $CV_{\mathrm{tr}}$ and age in the UV/blue bands, while the correlation in F814W is much weaker, although still statistically significant given the sample size.

Beyond the statistical significance, the binned trends in Fig.~\ref{fig:tr_veri_cmd}a indicate that the decline of $CV_{\rm tr}$ is not confined to the youngest age bin but persists across the full CMD-calibrated range. In the UV band, the median $CV_{\rm tr}$ decreases gradually with increasing age, without any abrupt transitions, supporting a broadly monotonic behaviour consistent with the non-parametric nature of the Spearman statistic. The interquartile ranges further show that the scatter in $CV_{\rm tr}$ remains substantial at all ages, but appears systematically larger at younger ages in the UV bands. This suggests that clusters at early evolutionary stages exhibit a wider diversity of internal structural complexity, while older clusters tend to occupy a narrower range of smoother morphologies. The presence of significant scatter also emphasizes that $CV_{\rm tr}$ is not expected to provide a deterministic age estimator for individual objects, but rather acts as a statistical structural proxy when applied to ensemble populations. 

To test the robustness of these correlations against sampling fluctuations, we performed bootstrap resampling (1000 realizations) of the cluster sample and recomputed the Spearman coefficient for each band. The resulting bootstrap statistics are summarized in Fig.~\ref{fig:tr_veri_cmd}b. In all bands, the bootstrap medians are consistent with the original measurements, and the 16th--84th percentile ranges remain narrow ($\Delta\rho \lesssim 0.05$ for F275W--F475W), demonstrating that the observed anti-correlations are not driven by a small subset of objects. Notably, the correlations in the UV/blue bands remain consistently negative across all bootstrap realizations, reinforcing the statistical robustness of the monotonic $CV_{\mathrm{tr}}$--age relation.

\begin{figure*}
    \centering
    \includegraphics[width=1 \linewidth]{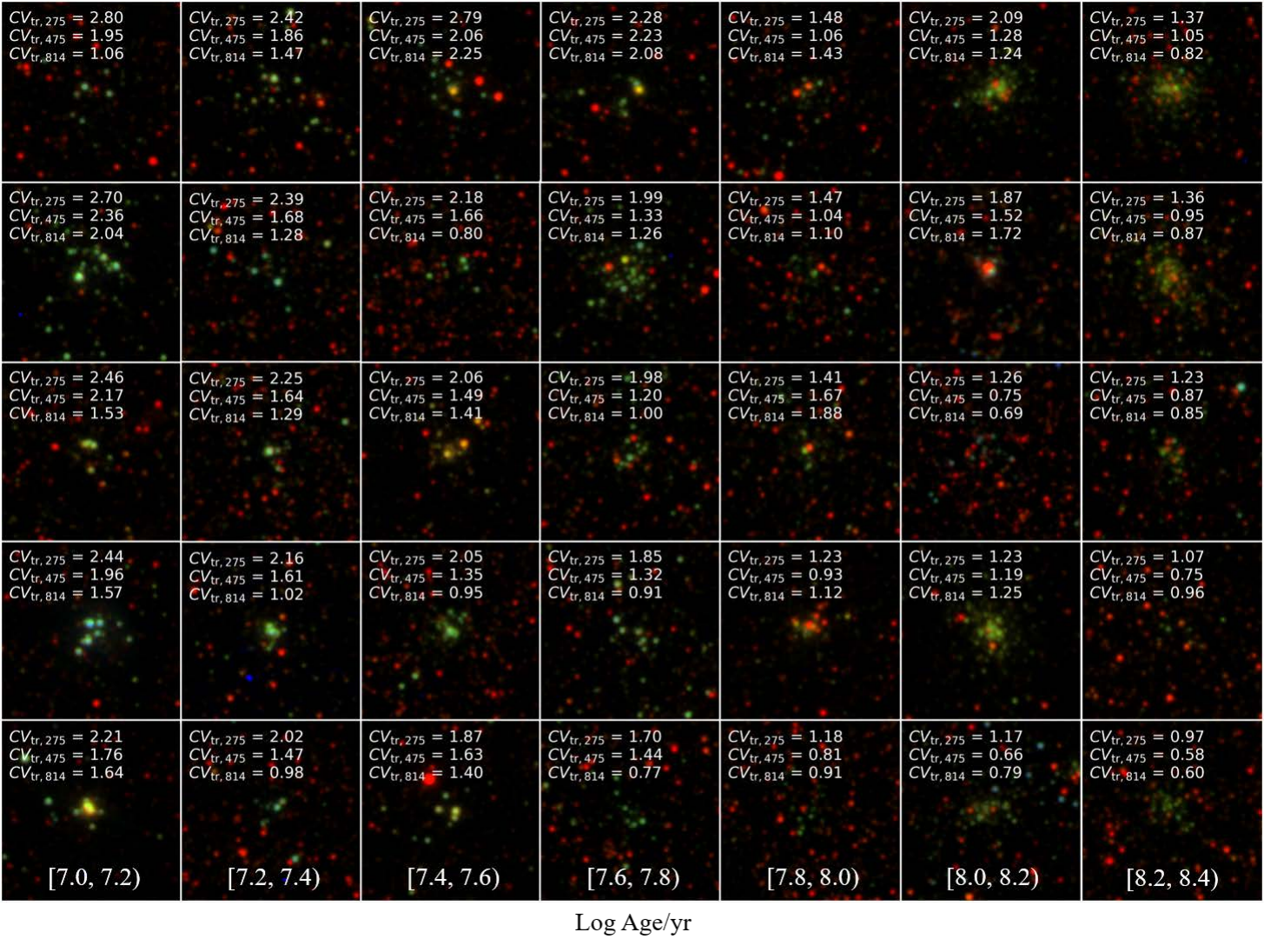}
    \caption{
    Three-colour cutouts of CMD-age-calibrated OBC candidates across the seven 0.2-dex age bins adopted in Figure~\ref{fig:tr_veri_cmd}. From left to right, the columns correspond to $\log_{10}(\mathrm{Age/yr})=[7.0,7.2), [7.2,7.4), \ldots, [8.2,8.4)$, with five clusters shown in each age bin. Within each column, the clusters are arranged from top to bottom in decreasing order of $CV_{\mathrm{tr},275}$. F814W, F475W, and F275W are mapped to the red, green, and blue channels, respectively. The labels in each panel give the measured $CV_{\mathrm{tr}}$ values in F275W, F475W, and F814W. Each cutout covers the same $6\arcsec \times 6\arcsec$ field centred on the cluster.}
    \label{fig:color_cutouts}
\end{figure*}

Figure~\ref{fig:color_cutouts} shows five illustrative clusters from each of the same 0.2-dex age bins. To construct the montage, the three clusters with the highest $CV_{\mathrm{tr},275}$ values were selected from each 0.1-dex age interval, and candidates from two adjacent intervals were combined to form each 0.2-dex bin. Of the resulting six candidates, the one with the highest $CV_{\mathrm{tr},275}$ was excluded to avoid representing each age bin by its most extreme case, leaving five clusters per bin. The corresponding $CV_{\mathrm{tr}}$ values measured in F275W, F475W, and F814W are indicated in each cutout.

The F814W, F475W, and F275W images were aligned to a common pixel grid, converted to $F_{\nu}$, and background-subtracted using sigma-clipped statistics before being mapped to the red, green, and blue channels, respectively. The three channels were combined using the RGB composition scheme of \citet{lupton2004preparing} implemented in \texttt{astropy.visualization.make\_lupton\_rgb}, adopting an asinh softening parameter of $Q=8$. The stretch parameter was determined separately from the flux distribution of each cutout to display both compact and diffuse structures effectively. The displayed colours qualitatively illustrate the multi-band morphology across the adopted age bins and are not intended to preserve photometric flux ratios.

\begin{table}
\centering
\caption{Linear fit parameters of $CV_{\mathrm{tr}}$ as a function of age in different bands.}
\label{tab:cv_age_fit}
\begin{tabular}{c c c}
\hline
Band & Intercept $b$ & Slope $k$ \\
\hline
F275W & 5.878 & -0.602 \\
F336W & 6.127 & -0.626 \\
F475W & 4.628 & -0.458 \\
F814W & 2.834 & -0.229 \\
\hline
\end{tabular}
\begin{flushleft}
\textit{Notes.} $b$ and $k$ represent the intercept and slope of the relation $CV_{\mathrm{tr}} = k_\lambda \cdot \log(\mathrm{Age/yr}) + b_\lambda$, respectively. 
\end{flushleft}
\end{table}

Subsequently, we performed a simple linear fit in each band using ordinary least squares (OLS) for all individual cluster measurements to quantify the overall $CV_{\mathrm{tr}}$--age trend:
\[
CV_{\mathrm{tr}} = k_\lambda \cdot \log(\mathrm{Age/yr}) + b_\lambda,
\]
where $b_\lambda$ and $k_\lambda$ are the intercept and slope for each band, respectively (see Table~\ref{tab:cv_age_fit}). The negative slopes across all bands indicate a systematic decrease of $CV_{\mathrm{tr}}$ with age. The steeper slopes in the UV and blue bands indicate that the age dependence of $CV_{\mathrm{tr}}$ is more pronounced at shorter wavelengths.

The fitted slopes in Table~\ref{tab:cv_age_fit} show that the negative $CV_{\mathrm{tr}}$--age relation is stronger in the UV/blue bands and becomes shallower towards F814W, consistent with the Spearman rank-correlation coefficients shown in Fig.~\ref{fig:tr_veri_cmd}. Over the age range covered by the CMD-calibrated sample, the stronger negative relation at shorter wavelengths may reflect the progressive fading of massive, UV-bright stars and the consequent decline in their contribution to the spatial inhomogeneity of the UV/blue light distribution. At redder wavelengths, however, the stochastic presence of luminous cool stars in short-lived evolved phases, such as red supergiants (RSGs), can introduce substantial cluster-to-cluster variations in the light distribution, thereby increasing the scatter and weakening the overall monotonic $CV_{\mathrm{tr}}$--age trend.

We caution that the CMD-calibrated sample does not fully cover the younger evolutionary phases, including those in which a small number of luminous RSGs may dominate the red-band light distribution and its second-order morphology. Therefore, the wavelength dependence of the $CV_{\mathrm{tr}}$--age relation should not be generalized to the full evolutionary sequence of star clusters. We emphasize, moreover, that the absolute normalization of $CV_{\mathrm{tr}}$ is band-dependent due to differing PSFs and noise properties; our interpretation therefore focuses on trends within each band and their relative strengths rather than on cross-band absolute values.

We further assess whether the observed $CV_{\rm tr}$--age relation could be driven by variations in cluster mass. Adopting the CMD-based mass estimates from \citet{johnson2016panchromatic}, which span $10^{2.5}$ to $10^{4}\,\mathrm{M_\odot}$, we compute Spearman rank correlations among $CV_{\rm tr}$, $\log t$, and $\log M$. We find a moderate positive correlation between age and mass ($\rho = 0.24$), indicating that older clusters in this subsample tend to be somewhat more massive. The correlation between $CV_{\rm tr}$ and mass is weaker, with $\rho = -0.24$ in F275W and F336W, $\rho = -0.12$ in F475W, and $\rho = 0.16$ in F814W. In comparison, the anti-correlation between $CV_{\rm tr}$ and age remains stronger in the UV/blue bands ($\rho \simeq -0.5$). Crucially, when we control for cluster mass using partial correlation analysis, the dependence of $CV_{\rm tr}$ on age remains robust. For instance, the correlation coefficient in F275W changes negligibly from $\rho = -0.54$ to $\rho_{\rm partial} = -0.51$, while in F814W, it shifts slightly from $\rho = -0.15$ to $-0.21$. This result confirms that the observed morphological evolution is intrinsically linked to cluster agieng and is not a secondary effect of cluster mass.

Taken together, these tests disfavour a purely mass-driven origin for the observed trend. These relative correlation strengths suggest that while cluster mass may introduce a secondary contribution to the observed structural variation, the dominant monotonic trend of $CV_{\rm tr}$ with age cannot be attributed solely to the underlying mass distribution within the CMD-calibrated sample.

\section{Discussion} \label{sec:discussion}

\subsection{Improvements and Limitations of the Current OBC Candidate Catalogue}
\label{sec:catalog_limitation}

The OBC candidate catalogue in M31 has been updated, and the number of candidates has increased from 578 to 747 due to an expanded spatial search grid, while preserving the search algorithm and overall spatial distribution pattern reported in previous studies \citep{liang2025comprehensive}. The algorithm is used primarily as a pre-selection step, with final candidate identification based on visual inspection. We also refine the definition of $R_{\mathrm{full}}$ using a non-parametric method based on the convergence of cumulative flux curves, which is better suited for candidates with irregular morphologies \citep[e.g.][]{ryon2017effective}. This reduces reliance on subjective visual estimation, though all boundaries are still visually reviewed. This refinement generally leads to more consistent $R_{\mathrm{eff}}$ values compared to nearby galaxy catalogues \citep{brown2021radii}.

However, even with the latest PHAST data (2025 release), only about two-thirds of M31 is covered, as shown in Fig.~2. The southwestern quadrant---particularly beyond 14~kpc and between $210^{\circ}$ and $330^{\circ}$ in azimuth---contains large gaps. We note that this incomplete coverage introduces substantial sampling bias, especially in dust-rich regions along the ring, limiting the statistical completeness of the catalogue. An additional limitation arises from the F275W-based candidate selection. Although the catalogue includes clusters with CMD-based ages extending to a few hundred Myr, the recovery of such objects does not imply completeness at these ages. As clusters evolve and their UV-bright massive-star populations fade, older systems become increasingly difficult to identify in F275W. The catalogue is therefore expected to become progressively incomplete toward the older end of the CMD-calibrated age range.

\subsection{Forward Modelling of Photometric Structural Evolution}\label{sec:phot_simu}

Motivated by the empirical wavelength-dependent trends in Section~\ref{sec:tensor_time}, we construct synthetic cluster images to demonstrate the plausibility of the observed $CV_{\rm tr}$-age trends under simplified but physically driven assumptions, and process them through the same measurement pipeline. 

The mock clusters are sampled over $\log(M/M_{\odot})=3.0$--4.0 in steps of 0.1~dex and $\log(\mathrm{age/yr})=6.6$--9.2 in steps of 0.2~dex. Internal substructure is represented by 1--3 fragmentation cores that are randomly placed within spatial extents of 1--4~pc, a range typical of young clusters' $R_{\rm full}$ in our sample.

Stellar masses are drawn from a Kroupa IMF \citep{kroupa2001variation} over $0.5$--$30\,M_{\odot}$. Individual stellar masses are sampled stochastically from the IMF until their cumulative mass reaches the prescribed cluster mass. The number of member stars is therefore not fixed a priori, but is determined by the cluster mass and each stochastic IMF realization.

The sampled member stars are then divided approximately equally among the fragmentation cores. Around each core centre, stellar positions are drawn from a three-dimensional Gaussian distribution with a dispersion of 0.5~pc along each Cartesian coordinate. This phenomenological prescription generates irregular, non-axisymmetric stellar distributions without imposing a smooth radial profile.

Multi-band luminosities are assigned using solar-metallicity MIST isochrones \citep{choi2016mesa,dotter2016mesa} in the same HST filters as the observations. The mock clusters are projected to the distance of M31 (0.78 Mpc) with random orientations. Synthetic images are generated with \texttt{GalSim} \citep{rowe2015galsim} by converting stellar fluxes to detector counts and convolving with instrument-specific PSFs (WFC3: STScI calibration products; ACS: empirical PSF from isolated stars; \citealt{sahu2021wfc3}). We further inject Gaussian noise with an amplitude of 0.5 times the median non-zero pixel value and repeat the analysis for ten noise realizations, finding $CV_{\rm tr}$--age relations indistinguishable from the noise-free case.

\begin{figure*}
    \centering
    \includegraphics[width=1 \linewidth]{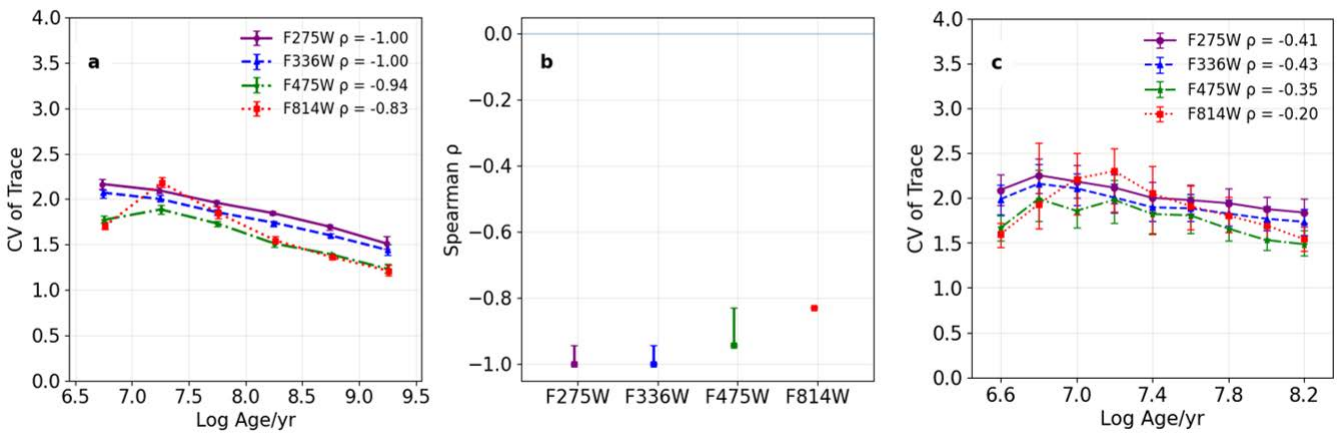}
    \caption{
    Evolution of the Hessian-based structure index $CV_{\rm tr}$ derived from synthetic cluster simulations.
    (a): Median $CV_{\rm tr}$ as a function of input cluster age for 38 independent mock realizations. Age is shown on a logarithmic scale, spanning $10^{6.6}$~yr to $10^{9.2}$~yr with a bin width of 0.5~dex. For each age, the symbol represents the median of the 38 realization-level median $CV_{\rm tr}$ values, and the vertical bars indicate their 16th--84th percentile range. The annotated $\rho$ values correspond to the median Spearman rank coefficients measured across all realizations.
    (b): Distribution of Spearman correlation coefficients obtained from the 38 independent simulations. Squares denote the median $\rho$ values, with vertical error bars representing the 16th--84th percentile intervals.
    (c): Same as panel (a), but for a more finely sampled young-age interval spanning $6.6 \leq \log(\mathrm{Age/yr}) \leq 8.2$ with a spacing of 0.2~dex. The median values and 16th--84th percentile ranges in this panel are calculated from all $CV_{\rm tr}$ measurements at each age across the 38 realizations, rather than from the realization-level medians. The annotated $\rho$ values are obtained in the same manner as in panel (a).}
    \label{fig:tr_veri_sim}
\end{figure*}

Figure~\ref{fig:tr_veri_sim} summarizes 38 independent mock realizations analysed with the same measurement procedure as the observations. Over the broad simulated age range (Figure~\ref{fig:tr_veri_sim}a), $CV_{\rm tr}$ shows an overall decline toward older ages in all four bands, although the evolution is not strictly monotonic at young ages. The Spearman coefficients $\rho$ calculated for the individual realizations remain consistently negative (Figure~\ref{fig:tr_veri_sim}b), with median values of $-1.00$, $-1.00$, $-0.94$, and $-0.83$ for F275W, F336W, F475W, and F814W, respectively, and with the 16th--84th percentile intervals remaining below zero. The relatively small realization-to-realization scatter compared with the overall dynamic range indicates that the general decline of $CV_{\rm tr}$ with age is robust against stochastic variations among the mock realizations.

Motivated by the possible influence of short-lived luminous stellar populations at young ages, we further examine the interval $6.6 \leq \log(\mathrm{Age/yr}) \leq 8.2$ using an age sampling of 0.2~dex (Figure~\ref{fig:tr_veri_sim}c). The finer sampling reveals additional non-monotonic structure that is not apparent in the more coarsely sampled trends. In both UV bands, the median $CV_{\rm tr}$ shows a modest enhancement around $\log(\mathrm{Age/yr})\simeq6.8$, followed by an overall decline toward older ages. The optical bands show stronger wavelength-dependent variations, with F814W exhibiting a more pronounced enhancement over $\log(\mathrm{Age/yr})\sim7.0$--$7.4$. Consequently, the rank correlations over this finely sampled young-age interval are weaker, with $\rho=-0.41$, $-0.43$, $-0.35$, and $-0.20$ for F275W, F336W, F475W, and F814W, respectively. These results show that the forward models reproduce an overall age-dependent evolution of $CV_{\rm tr}$, while also indicating local, wavelength-dependent departures from a simple monotonic relation at young ages. We therefore regard $CV_{\rm tr}$ as a statistical age-sensitive structural diagnostic rather than as a globally monotonic age indicator.

\begin{figure*}
    \centering
    \includegraphics[width=1 \linewidth]{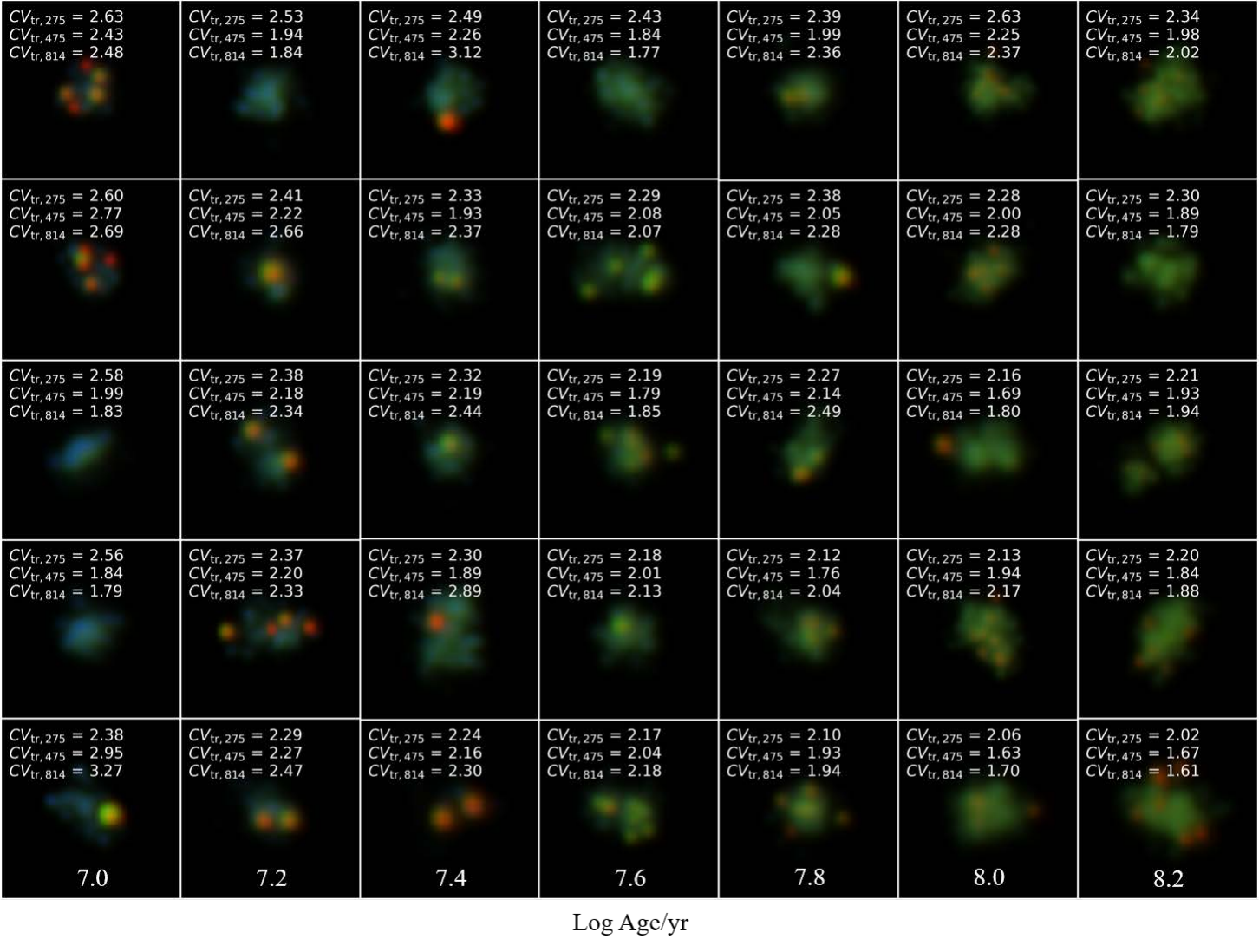}
    \caption{
    Three-colour cutouts of synthetic clusters at seven representative ages overlapping the age range of the observed OBC candidates. From left to right, the columns correspond to $\log_{10}(\mathrm{Age/yr})=7.0, 7.2, 7.4, \ldots, 8.2$, with five synthetic clusters shown at each age. Within each column, the clusters are arranged from top to bottom in decreasing order of $CV_{\mathrm{tr},275}$. F814W, F475W, and F275W are mapped to the red, green, and blue channels, respectively, after conversion to $F_{\nu}$, and the colour composites are constructed using the same procedure as adopted for the observed OBC candidates in Figure~\ref{fig:color_cutouts}. The labels in each panel give the measured $CV_{\mathrm{tr}}$ values in F275W, F475W, and F814W.}
    \label{fig:color_cutouts_sim}
\end{figure*}

To provide a direct visual comparison with the observed OBC candidates, Figure~\ref{fig:color_cutouts_sim} shows representative synthetic clusters over the age range shared with the CMD-calibrated observational sample. Five independent realizations are displayed at each of seven ages from $\log(\mathrm{Age/yr})=7.0$ to 8.2. The synthetic images are 
constructed using the same three-colour procedure as for the observed clusters, with F814W, F475W, and F275W mapped to the red, green, and blue channels, respectively. The montage illustrates the diversity of multi-band morphologies produced by the forward models while also showing the progressive reduction of high-contrast photometric substructure toward older ages, consistent with the quantitative $CV_{\rm tr}$ trends shown in Figure~\ref{fig:tr_veri_sim}.

We emphasize that the empirical validation presented in this work is constrained to the CMD-calibrated range ($\sim10$--300~Myr), while the extension to ages outside this range in the simulations is intended as an exploratory theoretical test rather than an observational claim. The clearer overall age dependence in the simulations likely reflects the controlled nature of the forward models and the absence of several real-world complications (e.g. crowding and complex backgrounds) that can dilute trends in the observational data. Accordingly, we interpret the simulations as a plausibility test for the emergence of age-dependent structural evolution, rather than as a quantitative match to the observed slope or scatter.

We further note that the forward models show an overall decline in $CV_{\rm tr}$ with age over the broad simulated age range, while the finer age sampling reveals local departures from strict monotonicity at young ages. These departures are wavelength dependent and are more pronounced in F814W. The observational data show a substantially weaker correlation in the red band. This difference likely reflects the wavelength-dependent sensitivity of $CV_{\rm tr}$ to stellar population evolution.

In the UV and blue bands, the light distribution is dominated by massive stars whose rapid fading and disappearance naturally enhance the evolution of light-weighted morphological contrast with age. The local variations revealed by the finer age sampling nevertheless indicate that this evolution need not be strictly monotonic at the youngest ages.

In the F814W band, while longer-lived intermediate- and low-mass stars contribute substantially to the red-band light, short-lived luminous post-main-sequence massive stars, such as red supergiants (RSGs), can also make important contributions in young stellar populations. The small number and rapid evolution of such luminous sources may introduce additional stochastic and age-dependent variations in the red-band light distribution, providing a plausible explanation for the more pronounced young-age structure in F814W. The present simplified models do not, however, isolate the contribution of RSGs explicitly. Their possible influence on $CV_{\rm tr}$ may offer an interesting avenue for using curvature-based diagnostics to probe short-lived luminous evolutionary phases in future work.

\section{Conclusion}\label{sec:conclusion}

In this work, we present a Hessian-based approach to quantify the internal photometric substructure of partially resolved OB cluster (OBC) candidates in M31, and assess its evolutionary relevance using independent CMD age estimates and forward modelling.

First, we refine and extend our previous OBC candidate sample \citep{liang2025comprehensive} using PHAT and PHAST imaging, yielding an updated catalogue of 747 candidates. The revised pipeline includes an expanded MeanShift search over a broader range of spatial scales and a semi-automated, growth-curve-based definition of the full-light radius $R_{\rm full}$, leading to more consistent size measurements across the survey footprint.

Second, we quantify internal photometric substructure via the Hessian-derived trace coefficient of variation, $CV_{\rm tr}$, measured consistently in four \textit{HST} bands. By cross-matching our catalogue with the subset of M31 clusters that have CMD-based ages from \citet{johnson2016panchromatic}, we identify 254 objects in common and find statistically significant anti-correlations between $CV_{\rm tr}$ and age in the UV/blue bands. This supports the interpretation of $CV_{\rm tr}$ as a structural proxy for relative evolutionary stage within the CMD-calibrated regime.

Finally, we construct synthetic cluster images spanning a broad range of ages and analyse them with the same measurement pipeline. The forward-modelling experiments robustly recover an overall decline of $CV_{\rm tr}$ with age over the broad simulated age baseline under simplified but physically motivated assumptions, while finer age sampling reveals local, wavelength-dependent departures from strict monotonicity at young ages. These departures are most pronounced in F814W, where the enhanced $CV_{\rm tr}$ at young ages is qualitatively consistent with a contribution from short-lived luminous evolved massive stars such as red supergiants (RSGs). These results support the interpretation that the observed age dependence of $CV_{\rm tr}$ can plausibly arise from stellar population evolution, while also showing that the relation need not be globally monotonic over the full age range. We emphasize that the empirical validation in this work is limited to the CMD-calibrated range ($\sim10$--300~Myr), while the extension to ages outside this range in the simulations is intended as a theoretical demonstration rather than an observational claim.

Future work will first focus on strengthening the empirical foundation of the curvature--age relation. In particular, incorporating more realistic treatments of observational noise, crowding, and instrumental effects will help to clarify the robustness of the observed trends. Extending the analysis to larger and more diverse cluster samples within M31 will also be important as additional homogeneous age estimates become available, including those derived from spectral energy distribution modelling (e.g. Prospector, \citealt{leja2017deriving, johnson2021stellar}). 

On the methodological side, further refinement of curvature-based diagnostics may improve sensitivity to internal structural evolution in the partially resolved regime. The exploration of alternative or higher-order Hessian-derived metrics, beyond the trace alone, could provide additional discriminatory power. We also note that the trace coefficient of variation is naturally sensitive to strong localized luminosity fluctuations. In the UV bands, where massive OB stars dominate the light budget, the presence of a small number of high-luminosity stars can enhance local curvature contrasts and broaden the distribution of $|\mathrm{Tr}(\mathbf{H})|$, potentially increasing $CV_{\rm tr}$. This suggests that, beyond its evolutionary sensitivity, the trace coefficient of variation may serve as a statistical indicator of clusters whose light is dominated by massive stars, particularly in the UV regime. An analysis combining spectroscopic classifications or resolved stellar content would be required to test this possibility in detail. 

In this context, a more systematic investigation of mass-dependent structural signatures will also be necessary. While the present analysis indicates that the curvature--age relation is not solely driven by cluster mass, the moderate correlation between mass and age suggests that mass-dependent sampling effects, particularly in the UV regime, may modulate the amplitude and scatter of $CV_{\rm tr}$. 
A study designed to disentangle stochastic massive-star sampling from secular structural evolution will therefore be pursued in future work.

Finally, the availability of larger cluster samples in nearby galaxies, enabled by forthcoming wide-field facilities such as the China Space Station Telescope (CSST; \citealt{zhan2021wide}), will allow curvature-based diagnostics to be tested across broader environmental conditions. A more comprehensive exploitation of the full Hessian matrix, including its eigenvalue structure and anisotropy properties, may ultimately contribute to a unified morphological framework for studying extragalactic star clusters beyond simple structural smoothing signatures.

\section*{Acknowledgements}

This work was supported by the National Natural Science Foundation of China (NSFC) under Grant Nos. 12588202, 12041302, and 12073038, the China Manned Space Engineering Program (China Space Station Telescope, CSST) under Grant Nos. CSST-2021-A08, CSST-2021-B02, CSST-2021-B03, and CSST-2021-B06, and the National Key R\&D Program of China under Grant No. 2023YFA1608004.

Data resources were provided by the China National Astronomical Data Center (NADC) and the Chinese Virtual Observatory (China‑VO). This research also benefited from the Astronomical Big Data Joint Research Center, co‑founded by the National Astronomical Observatories, Chinese Academy of Sciences, and Alibaba Cloud.

\section*{Data Availability}

The OBC candidate catalogues underlying this article are available at \url{https://nadc.china-vo.org/res/r101747/}. Additional processed files and scripts used in this study are available upon reasonable request from the corresponding author.



\bibliographystyle{mnras}
\bibliography{reference} 







\bsp	
\label{lastpage}
\end{CJK*}
\end{document}